\documentclass[11pt]{article}
\usepackage{graphicx}
\usepackage{amsfonts}

\usepackage{amsmath}
\usepackage{amsthm}
\usepackage{amssymb}
 
\usepackage[latin1]{inputenc}

\newcommand\be{\begin{equation}}
\newcommand\ee{\end{equation}}
\newcommand\bea{\begin{eqnarray}}
\newcommand\eea{\end{eqnarray}}

\newcommand\rrr{r}

\newcommand{\bp}{\mathbf{p}}
\newcommand{\bq}{\mathbf{q}}
\newcommand\RR{\mathbb{R}}
\newcommand{\dd}{{\mathrm d}}
\newcommand{\cM}{{\mathcal M}}
\newcommand{\De}{\Delta}
\newcommand\K{_{\rm K}} \newcommand\HO{_{\rm O}}
\newcommand{\ka}{\kappa}
\newcommand{\om}{\omega}
\newcommand{\rc}{{|\bq|_{\rm c}}}

\newcommand{\la}{\lambda}
\newcommand{\vp}{\varphi}
\newcommand{\pd}{\partial}

\newcommand\minus\backslash

\newcommand{\hH}{\hat{H}_\la}

\newcommand{\hbq}{\hat{\mathbf{q}}}
\newcommand{\hbp}{\hat{\mathbf{p}}}
\newcommand{\hC}{\hat{C}}
\newcommand{\hI}{\hat{C}}

\newcommand\rmi{{\rm i}}

\newcommand{\hq}{\hat{q}}
\newcommand{\hp}{\hat{p}}

\newtheorem{theorem}{Theorem}

\theoremstyle{definition}
\newtheorem{definition}[theorem]{Definition}

\newtheorem{example}[theorem]{Example}

\theoremstyle{remark}

\begin{document}

 \
 \smallskip
 \bigskip

 \bigskip

  \noindent
{\LARGE{\bf{New superintegrable models with position-dependent\\[4pt] mass from Bertrand's Theorem on curved spaces}}}

\bigskip

\bigskip

\bigskip

\begin{center}
{\large  \'Angel Ballesteros$^a$,   Alberto Enciso$^b$,  Francisco J. Herranz$^a$,\\[4pt] Orlando Ragnisco$^c$ and Danilo Riglioni$^c$}
\end{center}

\noindent
{$^a$ Departamento de F\1sica,  Universidad de Burgos,
09001 Burgos, Spain\\ ~~E-mail: angelb@ubu.es \quad fjherranz@ubu.es\\[10pt]
}
$^b$ Departamento de F\1sica Te\'orica II,   Universidad Complutense,   28040 Madrid,
Spain\\ ~~E-mail: aenciso@fis.ucm.es\\[10pt]
$^c$ Dipartimento di Fisica,   Università di Roma Tre and Istituto Nazionale di
Fisica Nucleare sezione di Roma Tre,  Via Vasca Navale 84,  00146 Roma, Italy  \\
~~E-mail: ragnisco@fis.uniroma3.it  \quad riglioni@fis.uniroma3.it    \\[10pt]

\medskip
\medskip

\bigskip

\begin{abstract} 
\noindent
 A generalized version of Bertrand's theorem on spherically symmetric curved spaces is presented. This result is based on the classification of $(3+1)$-dimensional (Lorentzian) Bertrand spacetimes, that gives rise to two families of Hamiltonian systems defined on certain 3-dimensional (Riemannian) spaces.  These two systems are shown to be either the Kepler or the oscillator potentials on the corresponding Bertrand spaces, and both of them are maximally superintegrable. Afterwards, the relationship between such  Bertrand Hamiltonians and position-dependent mass systems is explicitly established. These results are illustrated through 
the example of a superintegrable (nonlinear) oscillator on a Bertrand-Darboux space, whose quantization and physical features are also briefly addressed.
\end{abstract}

\newpage


\section{Introduction}

Bertrand's theorem, which dates back to the XIX century~\cite{Be73}, is a landmark
result in classical mechanics characterizing the Kepler and harmonic oscillator potentials in terms of their qualitative dynamics. More precisely~\cite{FK04}, Bertrand's theorem asserts that any spherically symmetric natural Hamiltonian system $H=\frac12|\bp|^2+ V(|\bq|)$ in (a subset of) $\RR^3$ that has a stable circular trajectory passing through each point in its configuration space and all whose bounded trajectories are closed is either a harmonic oscillator ($V(r)=A\,r^2+B$) or a Kepler system $(V(r)=A/r+B$).

Surprisingly, the classical theorem of Bertrand found a natural extension in and application to general relativity some fifteen years ago thanks to a remarkable paper of Perlick~\cite{Pe92}. Indeed, the author undertook the classification of all {\em Bertrand spacetimes}, which, roughly speaking, are spherically symmetric and static spacetimes whose timelike geodesics satisfy properties analogous to those of the trajectories of the harmonic oscillator or Kepler systems. The connection between timelike geodesics in spacetime and the trajectories of a classical Hamiltonian system is that, if one writes the Lorentzian metric as 
$$
\eta=g_{ij}(q)\,\dd q^i\,\dd q^j-V^{-1}(q)\,\dd t^2 ,
$$
where $g$ is a Riemannian metric on a $3$-manifold, the timelike geodesics in spacetime are naturally related to the trajectories of the Hamiltonian system 
$$
H=g^{ij}(q)p_ip_j+V(q).
$$

Perlick's classification of Bertrand spacetimes consisted of two multi-parametric families. When one additionally imposes that the spacial part of the metric be Euclidean, Perlick's result becomes tantamount to the classical Bertrand's  theorem. Moreover, the general case includes a number of other systems that have received considerable attention, particularly in connection with integrable monopole motion and the existence of generalized Runge--Lenz vectors. For instance, Perlick's classification includes spacetimes constructed over the $3$-sphere, the hyperbolic $3$-space and the Iwai--Katayama spaces~\cite{IK94,IK95}, which generalize the Taub--NUT spacetime.

In this paper we   review several recent results related to this problem and we also establish  a natural connection between curved Bertrand systems and position-dependent mass (PDM) Hamiltonians~\cite{Roos}--\cite{Levai}.  We   begin by showing, in the next section, that the rather complicated families of Bertrand spacetimes admit a strikingly simple physical interpretation~\cite{BEHR08}: they correspond to either an intrinsic oscillator or an intrinsic Kepler system. Next we   address  in section~\ref{S.Bertrand} the superintegrability of the associated Hamiltonian system and recall the construction of additional integrals of motion of Runge--Lenz type, which settled in a satisfactory way a problem with a large body of related literature (cf.~\cite{BEHR09, commun2} and references therein). This allows us to state an optimal version of Bertrand's theorem on Riemannian manifolds.  Furthermore, we present in section 4 the application of the above results to PDM Hamiltonians by rewriting the previous Bertrand Hamiltonians in terms of a variable mass function. 
All of these results are explicitly illustrated by       discussing  in section 5  the $N$-dimensional ($N$D) version of one of the most interesting Bertrand Hamiltonians: the   Darboux III Hamiltonian~\cite{BEHR08a}. The associated quantum mechanical problem  together with some possible physical applications of these new PDM integrable models are also considered in the last section.


\section{Bertrand spacetimes}

To begin with, let us recall Perlick's definition of a Bertrand spacetime~\cite{Pe92}. We will consider a spherically symmetric, static spacetime $(\cM\times\RR,\eta)$, where $\cM$ is a $3$-manifold. This ensures that the Lorentzian metric $\eta$ can be written as
$$
\eta=h(r)^2\,\dd r^2+r^2\big(\dd\theta^2+\sin^2\theta\,\dd\varphi^2\big)-\frac{\dd t^2}{V(r)}\,,
$$
where $V$ is a smooth scalar function and
\begin{equation}\label{metric}
  g=h(r)^2\,\dd r^2+r^2(\dd\theta^2+\sin^2\theta\,\dd\varphi^2)
\end{equation}
defines a Riemannian metric on $\cM$.
Following Perlick, by a {\em trajectory} in spacetime we mean the projection of an inextendible timelike geodesic to a (fixed but otherwise arbitrary) constant time leaf $\cM×\{t_0\}$. This terminology is motivated by the fact that a trajectory in spacetime actually corresponds to a trajectory (in configuration space) of the Hamiltonian
\begin{equation}\label{Hamiltonian}
  H=\frac12g^{ij}p_ip_j+V=\frac12\Bigg(\frac{p_r^2}{h(r)^2}
+\frac{p_\theta^2}{r^2}+\frac{p_\varphi^2}{r^2\sin^2\theta}\Bigg)+V(r)
\end{equation}
in $\cM$. As customary, $g^{ij}$ is the inverse matrix to $g_{ij}$, $p\in T^*\cM$ is the momentum and $(p_r,p_\theta,p_\varphi)$ are the conjugate momenta of the coordinates $(r,\theta,\varphi)$.

\begin{definition}\label{D.Bertrand}
The Lorentzian $4$-manifold $(\cM\times \RR,\eta)$ is a {\em Bertrand spacetime} if:
\begin{enumerate}
\item There is a circular ($r=\text{const.}$) trajectory passing through each point of $\cM$.

\item The above circular trajectories are stable, that is, any initial condition sufficiently close to that of a circular trajectory gives a periodic trajectory.
\end{enumerate}
\end{definition}

Perlick's classification of all Bertrand spacetimes~\cite{Pe92} can then be stated as follows:

\begin{theorem}[Perlick]\label{T.Perlick}
  The metric of a Bertrand spacetime can be expressed in exactly one of the following forms:
  \begin{enumerate}
  \item Type I:
    \[
g=\frac{m^2\dd r^2 }{n^2\left( 1+Kr^2\right) }
+r^2(\dd\theta^2+\sin^2\theta\,\dd\varphi^2)\,,\qquad V=\sqrt{ r^{-2} +K} +G\,.
\]
\item Type II:
  \begin{align*}
g&=\frac{2m^2\left( 1-Dr^2\pm\sqrt{(1-Dr^2)^2-Kr^4}
\right)}{n^2\left((1-Dr^2)^2-Kr^4
\right)}\,\dd r^2 +r^2(\dd\theta^2+\sin^2\theta\,\dd\varphi^2)\,,\\ V &=G\mp r^2\Big(1-Dr^2\pm\sqrt{(1-Dr^2)^2-Kr^4}\Big)^{-1} .
\end{align*}
\end{enumerate}
Here $m,n$ are coprime nonnegative integers and $D,G,K$ are real constants.    
\end{theorem}

Notice that in~\cite{Pe92, BEHR08} this result was expressed through the rational number $\beta=n/m$.

To provide an interpretation of the above (rather involved) formulas for the Bertrand spacetimes, it is convenient to define the concept of harmonic oscillator and Kepler potential in any spherically symmetric $3$-manifold. To this end, let us use the notation~\eqref{metric} for a spherically symmetric metric on $\cM$ and consider its associated Laplacian, which we denote by $\De_g$. It is standard that if $u(r)$ is function on $\cM$ that only depends on the radial coordinate, then its Laplacian is also radial and given by
\[
\De_gu(r)=\frac{1}{r^2h(r)}\frac{\dd}{\dd r}\bigg(\frac{r^2}{h(r)}\frac{\dd u}{\dd r}\bigg)\,.
\]
Then the symmetric Green function $u(r)$ is obtained as  the solution of the equation $\De_gu(r)=0$ on $\cM\backslash \{\bf 0\}$, namely  
\be
u(r)=\int^r \frac{h(r')}{r'^2}\,\dd r' .
\label{green}
\ee
As the Kepler potential in 3D Euclidean space is simply the radial Green function $u(r)$ of the Laplacian and the harmonic oscillator is its inverse square, it is natural to make the following

\begin{definition}\label{D.KHO}
The {\em Kepler} and the {\em harmonic oscillator potentials} in $(\cM,g)$ are respectively given by the radial functions
$$
V\K(r)=A_1\bigg(\int^r_ar'^{-2}h(r')\,\dd r'+B_1\bigg)\,,\qquad V\HO(r)=A_2\bigg(\int^r_ar'^{-2}h(r')\,\dd r'+B_2\bigg)^{-2}\,,
$$
where $a,A_j,B_j$ $(j=1,2)$ are constants.
\end{definition}

This definition is obviously valid in higher dimensions as well.

\begin{example}\label{E:CCS}
Let $(\cM,g)$ be the simply connected  3D space   of constant sectional curvature $\ka$. In this case the metric has the form~\eqref{metric} with $h(r)^2=1/({1-\ka r^2})$, so that the corresponding Kepler and harmonic oscillator potentials are 
$$
V\K=\sqrt{r^{-2}-\ka}\,,\qquad V\HO=({r^{-2}-\ka})^{-1} ,
$$
up to additive and multiplicative constants. In terms of the distance function $\rho_\ka$ to the point $r=0$ these can be rewritten as
\[
V\K=\sqrt\ka\cot\big(\sqrt\ka\,\rho_\ka\big)\,,\qquad V\HO= {\tan^2(\sqrt\ka\,\rho_\ka)} / \ka  ,
\]
thus reproducing the known prescriptions for the sphere and the hyperbolic space~\cite{BH07,Sh05}. \end{example}

This readily gives the following interpretation of the Bertrand spacetimes~\cite{BEHR08}:

\begin{theorem}\label{T.potentials}
In a type I (resp.\ type II) Bertrand spacetime, $V$ is the intrinsic Kepler (resp.\ harmonic oscillator) potential associated with $g$.
\end{theorem}


\section{Bertrand's theorem in 3D curved spaces}
\label{S.Bertrand}

Let us now consider the maximal superintegrability (MS) of the Bertrand Hamiltonians~\eqref{Hamiltonian}. It is well known that, the Bertrand Hamiltonians being spherically symmetric, to establish their MS it suffices to obtain a functionally independent additional integral. As it turns out, the most convenient way to obtain this additional integral is as a `generalized Runge--Lenz tensor'.

Indeed, Bertrand Hamiltonians are somehow similar to the usual harmonic oscillator and Kepler systems in Euclidean space. It is classical that the MS of the Kepler system can be readily proved using that the Runge--Lenz vector is conserved~\cite{GS90}. In the case of the harmonic oscillator, there is no natural way of defining a conserved Runge--Lenz {\em vector}, but the elements of the symmetric {\em Fradkin matrix} $C_{ij}=2\om^2q_iq_j+p_ip_j$ are constants of the motion and encode the main algebraic properties of the model~\cite{Fr65}. A straightforward computation shows that the standard Kepler system is obtained from the type I Bertrand Hamiltonian by setting $K=0$ and $n=m=1$, while the harmonic oscillator is the type II Hamiltonian with parameters $K=D=0$, $n=2$ and $m=1$. The statement of the following key theorem, which proves the MS of Bertrand Hamiltonians~\cite{BEHR09}, is therefore not surprising:

\begin{theorem}\label{T.superint}
  There exists a (nontrivial) rank-$n$ symmetric tensor field invariant under the flow of the Bertrand Hamiltonian, where $n$ is the parameter introduced in theorem~\ref{T.Perlick}.
\end{theorem}

The proof of this theorem has been fully discussed in~\cite{BEHR09}. Here we want to stress that the parameter $n$ plays a crucial role in the construction of the first integrals. The basic observation, which goes back to Fradkin~\cite{Fr67}, is that any spherically symmetric Hamiltonian $H_0=\frac12|\bp|^2+U(|\bq|)$ preserves the unit vector field
$$
{\mathbf a}=\frac{\cos\vp}r\,\bq+\frac{\sin\vp}{rJ}\,\bq \times(\bq\times\bp)\,, \qquad J=p_\varphi=r^2\dot\varphi. 
$$
Of course, this only shows  the existence of a {\em local} additional integral, which is trivial in view of the flow-box theorem, but the point is that this provides a bona fide global first integral provided one can express $\cos\vp$ and $\sin\vp/J$ in terms of $\bp$ and $\bq$. In the case of the Kepler problem, this readily yields a conserved vector field which is essentially the Runge--Lenz vector divided by its norm. In the case of the harmonic oscillator,   the above quantities are not well defined functions of $\bp$ and $\bq$, but they do define an analogous bi-valued conserved vector field. By taking the two-fold symmetrized tensor product of this vector field one can remove this indeterminacy, and this procedure yields an invariant $2$-tensor essentially analogous to the conserved matrix $C_{ij}$ mentioned above. The proof of theorem~\ref{T.superint} follows this line of thought a bit further. Ultimately, the role $n$ plays in the proof merely reflects the properties of the trajectories of the Hamiltonian (studied in~\cite{Pe92,BEHR08}), which depend crucially on the numbers $n$ and $m$. It should be noticed that the dependence of the constants of motion on the momenta could be extremely complicated, which is the reason why only a few among the Bertrand Hamiltonians had previously been identified as maximally superintegrable systems.

To summarize, let us state the complete version of the optimal extension of Bertrand's theorem to Riemannian manifolds~\cite{BEHR09}:

\begin{theorem}\label{T.}
Let $H=\frac12g^{ij}p_ip_j+V$ be a Bertrand Hamiltonian, i.e., a spherically symmetric, natural Hamiltonian system on a Riemannian 3-manifold $(\cM,g)$ that has a stable circular trajectory passing through each point in its configuration space and whose bounded trajectories are all closed. Then the following statements hold:
\begin{enumerate}
\item The metric $g$ and the potential $V$ are of the form given in theorem~\ref{T.Perlick} for some coprime positive integers $n,m$.

\item The potential is the intrinsic Kepler or the harmonic oscillator potential of $(\cM,g)$.

\item $H$ is superintegrable. More precisely, there exists a nontrivial rank $n$ tensor field which is invariant and plays the role of the Runge--Lenz vector.
\end{enumerate}
\end{theorem}


\section{Curved Bertrand systems as  classical PDM Hamiltonians}

So far, by starting from Perlick's classification of $(3+1)$D Bertrand spacetimes, we have obtained two families of MS Hamiltonians on 3D    Riemannian manifolds, which are either  of Kepler or oscillator type. These results can also be translated into the language of PDM systems~\cite{Roos}--\cite{Levai}. For this purpose, we remark that Bertrand Hamiltonians are initially  expressed in the form (\ref{Hamiltonian}), that is,  $H=\frac12g^{ij}p_ip_j+V(r)$, with the potential $V$ being determined by theorem 2. Hence it is necessary to rewrite the
Hamiltonian in  terms of a variable mass $M(|\bq|)$  in  the form
\be
H=\frac{ \bp^2 }{ 2M(|\bq|) }+V(|\bq|) .
\label{pdm}
\ee
At this point we stress that the `radial' Bertrand coordinate $r$ is by no means $|\bq|$. Therefore, the translation has to be achieved by    defining the appropriate change 
of coordinates   $r\leftrightarrow|\bq|$ in the underlying metric (\ref{metric}) of the Bertrand Hamiltonians, thus giving rise to a conformally flat metric:
\be
 g=h(r)^2\,\dd r^2+r^2(\dd\theta^2+\sin^2\theta\,\dd\varphi^2)=f(|\bq|)^2\,\dd\bq^2.
 \label{metric3}
\ee
This yields the relations
\be
r=|\bq| f(|\bq|) ,\qquad f(|\bq|)\dd|\bq|=h(r)\dd r ,\qquad \frac{1}{|\bq|}\,\dd|\bq|=\frac{h(r)}{r}\,\dd r .
\label{relations}
\ee
Consequently, the variable mass can thus be  related to the conformal factor by setting 
\be
M(|\bq|)= m_0  f(|\bq|)^2=\frac{m_0 r^2}{\bq^2},
\label{mass}
\ee
where $m_0$ is a positive real constant that hereafter we shall fix to 1.
In the following we present  this relationship for each of the two types of Bertrand Hamiltonians.


\subsection{Type I: Bertrand-Kepler Hamiltonians}

In this case, by taking into account that
$$
h(r)=\frac{m}{n\sqrt{1+K r^2}}
$$
and by applying relations (\ref{relations}) we obtain that
$$
 |\bq|=\left( \frac{r}{1+\sqrt{1+K r^2}}\right)^{m/n} ,\qquad  r=\frac{2}{|\bq|^{-(n/m)}-K |\bq|^{(n/m)}} .
$$
Therefore the position-dependent mass function is given by
\be
M(|\bq|)=\frac 4{ \left( |\bq|^{-(n/m)}-K |\bq|^{(n/m)} \right)^{2}\bq^2 },
\label{mass1}
\ee
and the resulting PDM Hamiltonian provided by theorem 2 turns out to be 
\be
H=\frac 18 \left( |\bq|^{-(n/m)}-K |\bq|^{(n/m)} \right)^2 \bq^2\,  \bp^2+ A_1 \left\{K+\frac 14   \left( |\bq|^{-(n/m)}-K |\bq|^{(n/m)} \right)^2\right\}^{1/2}.
\label{pdm1}
\ee

\begin{example}    Riemannian spaces of constant  sectional curvature $\ka$ arise in the Bertrand Hamiltonians of type I 
when $n=m=1$ and $K=-\ka$~\cite{BEHR08}. Hence the variable mass (\ref{mass1}) and the Hamiltonian (\ref{pdm1}) reduce to
$$
M(|\bq|)=\frac{4}{( 1+\ka \bq^2   )^{2}},\qquad    H_\ka=\frac 18 \left( 1+\ka \bq^2  \right)^2   \bp^2+ A_1\frac{1-\ka \bq^2}{2|\bq|} ,
\label{ex8}
$$
which is the known  Kepler system written in Poincar\'e coordinates~\cite{BH07,kepler} on the spherical $(\ka>0)$, hyperbolic  $(\ka<0)$ and  Euclidean $(\ka=0)$ spaces.  Clearly, we can scale the Hamiltonian as $4 H_\ka$. The corresponding Runge--Lenz vector in these coordinates can be found in~\cite{BH07,kepler}.
\end{example}


\subsection{Type II: Bertrand-oscillator Hamiltonians}
 
Now   the function $h(\rrr)$  is given by
\be
h(r)=\frac{m\sqrt{2}}{n}\left(\frac{ 1-D\rrr^2\pm\sqrt{(1-D\rrr^2)^2-K\rrr^4}  
 }{ (1-D\rrr^2)^2-K\rrr^4  }\right)^{1/2} ,
\label{oa}
\ee
 and the relation between  the variables $|\bq|$ and $r$ defined through (\ref{relations})   can be obtained as
 \be
|\bq|=\exp\left\{ r\, u(r)-\int^r u(r')\dd r' \right\} ,
\label{ooaa}
\ee
(up to an additive constant coming from   the integral) where $u(r)$ is the Green function (\ref{green}). For these type II systems $u(r)$ reads as
$$
u(\rrr)= \mp \frac{m\sqrt{2}} {n\rrr} \left( 1-D\rrr^2\pm\sqrt{(1-D\rrr^2)^2-K\rrr^4}
\right)^{1/2} .
\label{oc}
$$
Therefore,  the explicit   general  result  for $M(|\bq|)=r^2/\bq^2$ is  quite cumbersome and we omit it here. However, for some particular cases it adopts a simple form, as it is shown in the sequel.

\begin{example}  The three Riemannian spaces of constant    curvature $\ka$  now appear by setting 
  $n=2$, $m=1$, $K=0$, $D=\ka$ and by taking the  positive sign   within the two posibilities `$\pm$' in (\ref{oa})~\cite{BEHR08}. 
  In this way we find that
  $$
  h(r)=\frac{1}{\sqrt{1-\ka r^2}},\qquad u(r)=-\frac{\sqrt{1-\ka r^2}}{r} ,
  $$
  so that
  $$
 r= \frac{2|\bq|}{1+\ka \bq^2},\qquad  |\bq|=\frac{r}{1+\sqrt{1-\ka r^2}},
  $$
  provided that we have dropped an additive constant $\ln 2$ in the integral (\ref{ooaa}). Then, as expected, we obtain the same variable mass as in the previous example and    the corresponding Bertrand oscillator in Poincar\'e variables is given by
  $$
   H_\ka=\frac 18 \left( 1+\ka \bq^2  \right)^2   \bp^2+ A_2\frac{2\bq^2}{(1-\ka \bq^2)^2} .
$$
  The   integrals $C_{ii}\equiv I_i$ in the diagonal of the conserved matrix $C_{ij}$ can be found in~\cite{BH07}. 
\end{example}


\section{The   Darboux III oscillator}

In what follows we focus our attention on a particular system of the family of Bertrand  Hamiltonians of type II, the so called 
Darboux III oscillator. The underlying Bertrand space is the 3D version of the Darboux surface of type III~\cite{Ko72, KKMW03}, for which an $N$D spherically symmetric generalization   was constructed in~\cite{PLB,annals}. Such a 3D Darboux-Bertrand space 
 corresponds to choose the $+$ sign in (\ref{oa}) and to set $n=2$,  $m=1$,  $K=D^2$ and $D=-2\la$, where $\la$ is a real parameter. This yields
 $$
 h(r)=\frac{1+\sqrt{1+4\la r^2}}{2\sqrt{1+4\la r^2}} ,\qquad u(r)=-\left(\frac{1+ \sqrt{1+4\la r^2}}{2 r} \right) .
 $$
Therefore the transformations between the radial variables $r$ and $|\bq|$ turn out to be
$$
r=|\bq| \sqrt{1+\la \bq^2},\qquad |\bq|=\left( \frac{\sqrt{1+4\la r^2}-1}{2\la}\right)^{1/2} .
$$
Then the variable mass function reads $M(|\bq|)=1+\la \bq^2$ and the resulting Darboux Hamiltonian is given by
\be
H_\la= \frac{\bp^2}{2(1+\la \bq^2)} +  \frac{ \om^2 \bq^2}{2(1+\la \bq^2)},
\label{ac}
\ee
where we have written $A_2=\om^2/2$. According to section 3, the 3D Hamiltonian $H_\la$ is a MS system, since it is endowed with a conserved matrix $C_{ij}$ (a curved Fradkin tensor). In fact, this result can directly be extended to arbitrary dimension $N$, as it has been proven in~\cite{BEHR08a,gadella}:

\begin{theorem} 
(i) The Hamiltonian ${  H}_\la$  (\ref{ac}),  for any dimension N  and for any real value of $\la$, is endowed with the following constants of motion.

\noindent
$\bullet$ $(2N-3)$  angular momentum integrals:
\be
  C^{(m)}=\!\! \sum_{1\leq i<j\leq m} \!\!\!\! (q_ip_j-q_jp_i)^2 , \qquad 
 C_{(m)}=\!\!\! \sum_{N-m<i<j\leq N}\!\!\!\!\!\!  (q_ip_j-q_jp_i)^2 ,   \label{af}
 \ee
 where $m=2,\dots,N$ and $C^{(N)}=C_{(N)}$.
 
 \noindent
$\bullet$ $N^2$ integrals  given by the components of the ND curved Fradkin tensor:
 \be
 C_{ij}=p_ip_j-\bigl(2\la  {H}_\la(\bq,\bp)-\om^2\bigr) q_iq_j , 
\label{ag}
\ee
where $ i,j=1,\dots,N$ and 
such that    ${  H}_\la=\frac 12 \sum_{i=1}^N C_{ii}$.  

\noindent
(ii) Each of the three  sets $\{{  H}_\la,C^{(m)}\}$,  
$\{{  H}_\la,C_{(m)}\}$ ($m=2,\dots,N$) and   $\{C_{ii}\}$ ($i=1,\dots,N$) is  formed by $N$ functionally independent functions  in involution.

\noindent
(iii) The set $\{ {  H}_\la,C^{(m)}, C_{(m)},  C_{ii} \}$ for $m=2,\dots,N$ with a fixed index $i$    is  constituted  by $2N-1$ functionally independent functions. 
\end{theorem}

We remark that the constants of motion (\ref{af}) and (\ref{ag}) can also be obtained~\cite{gadella} from   the  {\em free Euclidean  motion} by means of a St\"ackel transform or coupling constant metamorphosis (see~\cite{Hietarinta,Stackel5} and references therein).

It is also worth stressing that although the above statement holds for any real value of $\la$, the specific resulting system does depend on such a value, in such a manner that $H_\la$ comprises, in fact, {\em three} different nonlinear physical systems~\cite{gadella}:

\begin{itemize}

\item {\em Nonlinear hyperbolic oscillator}. For $\la>0$ the Darboux space is the complete Riemannian manifold $\cM^N=(\RR^N,g)$ with metric  $g_{ij}=(1+\la\bq^2)\,\delta_{ij}$.  The    scalar curvature $R(|\bq|)$  has a minimum at the origin $R(0)=-2\la N(N-1)$,  which coincides with the scalar curvature of the $N$D {\em hyperbolic space} with negative constant sectional curvature  $\ka=-2\la$.

\item {\em Nonlinear  spherical  oscillator}. For $\la<0$ we firstly consider   the interior   Darboux space defined by $\cM^N=(B_{\rc},g)$, 
where $g_{ij}=(1-|\la|\bq^2)\,\delta_{ij}$ and $B_\rc$ denotes the ball centered at $0$ of radius $\rc=1/\sqrt{|\la|}$ (the critical   value for the metric and for $H_\la$).  Now  $R(0)=2|\la| N(N-1)$ is exactly   the    scalar curvature of the $N$D {\em spherical space} with positive constant sectional curvature  $\ka=+2|\la|$.

 \item {\em Nonlinear  exterior  potential}. For $\la<0$ we can also consider   the exterior   Darboux space defined by    $\cM^N=(\RR^N\minus\overline{B_{\rc}},g)$; this implies to reverse the sign of   the metric and, therefore, of the Hamiltonian itself, namely,
 $g_{ij}=(|\la|\bq^2-1)\,\delta_{ij}$. This system can naturally be interpreted as an infinite barrier potential rather that an oscillator one.

\end{itemize}

\newpage


\section{Quantization of the Darboux III oscillator}

The quantization problem for Bertrand Hamiltonians arises as a challenging research program, whose first steps in the case of the Darboux III oscillator are summarized as follows.
 
 Let us consider 
the quantum Cartesian coordinates and momenta, $\hbq$, $\hbp$, with Lie brackets and   differential representation given by
$$
[\hq_i,\hp_j]=\rmi \hbar \delta_{ij},\qquad \hq_i=q_i,\qquad \hp_i=-\rmi  \hbar \partial_i=-\rmi  \hbar \frac{\partial}{\partial q_i},
\qquad  \De=\frac{\pd^2}{\pd^2 q_1}+\cdots +\frac{\pd^2}{\pd^2 q_N}.
\label{ka}
$$
 Our aim now is to    construct the quantum mechanical counterpart of the $N$D classical Hamiltonian (\ref{ac}): ${  H}_\la(\bq,\bp)\to \hat{  H}_\la(\hat\bq,\hat\bp)$. As it is well known,  the crucial point   is  to obtain the quantum analogue of the kinetic term, since there is an order ambiguity in its quantization.  This task can be faced by applying three different quantization procedures~\cite{future}: (i)  the  `Schr\"odinger quantization'; (ii)   the Laplace--Beltrami   quantization (which makes use of the   Laplace operator on curved spaces); and (iii)   a  PDM quantization. 

We stress that  if we impose that the quantum Hamiltonian  $\hat{  H}_\la$  keeps the maximal superintegrability (that is, the existence of $2N-2$ algebraically independent operators that commute with ${ H}_\la$),  then only the Schr\"odinger quantization yields, in a direct way, to fulfill this condition. Nevertheless, the Laplace--Beltrami   and PDM quantizations also lead to  MS quantum Hamiltonians once an additional `pure' quantum potential term is added to   the initial quantum Hamiltonian, and such     potential terms are related through   gauge transformations  to the  Schr\"odinger quantization. The resulting  MS  Schr\"odinger  quantization of $H_\la$ (\ref{ac}) is   characterized   as follows~\cite{FTC} (this result is worth to be compared with theorem 10).

  \begin{theorem}
    Let   $\hH$ be the ND quantum Bertrand-Darboux Hamiltonian given by
 \be
 {\hH}= 
 \frac{1}{2(1+\la \hbq^2)}\, \hbp^2+ \  \frac{ \om^2 \hbq^2}{2(1+\la \hbq^2)} =\frac{1}{2(1+\la \bq^2)}\big(-\hbar^2\De+\om^2\bq^2\big).
 \label{ca}
 \ee
 For any real value of $\la$ the following statements hold:
 
\noindent
(i) $\hH$ commutes with the following observables:

\noindent
$\bullet$ $(2N-3)$ quantum angular momentum operators,
\be
  \hC^{(m)}=\!\! \sum_{1\leq i<j\leq m} \!\!\!\! ( \hq_i \hp_j- \hq_j \hp_i)^2  , \qquad 
  \hC_{(m)}=\!\!\! \sum_{N-m<i<j\leq N}\!\!\!\!\!\!  ( \hq_i \hp_j- \hq_j \hp_i)^2  ,  \label{cb}
 \ee
where $m=2.\dots,N$ and    $\hC^{(N)}=\hC_{(N)}$.

\noindent
$\bullet$ $N^2$ operators which form an ND quantum Fradkin tensor,
  \be
  \hC_{ij}= \hp_i\hp_j- 2\la  \hq_i \hq_j{ \hH}( \hbq, \hbp)+ \om^2 \hq_i\hq_j , 
\label{cc}
\ee
where $i,j=1,\dots,N$ and   such that    ${\hH}=\frac 12 \sum_{i=1}^N \hC_{ii}$.  

\noindent
(ii) Each of the three  sets $\{{\hH},\hC^{(m)}\}$,  
$\{{\hH},\hC_{(m)}\}$ ($m=2,\dots,N$) and   $\{\hI_{ii}\}$ ($i=1,\dots,N$) is  formed by $N$ algebraically  independent  commuting observables.
\smallskip\\
(iii) The set $\{ { \hH},\hC^{(m)}, \hC_{(m)},  \hI_{ii} \}$ for $m=2,\dots,N$ with a fixed index $i$    is  formed by $2N-1$ algebraically independent observables. \\
(iv) $\hH$ is formally self-adjoint on the Hilbert space $L^2(\cM^N)$, endowed with the scalar product
\[
\langle \Psi | \Phi \rangle_\la = \int_{\cM^N} \overline{{\Psi}(\bq)} \Phi(\bq)(1+\la\bq^2){\dd}\bq .\]
\end{theorem}

Clearly, the results of  theorem 11 should be adapted to each of the three different systems described in the previous section. In particular, we consider here the quantum hyperbolic-type Hamiltonian,  that has been fully solved in~\cite{FTC}.


\subsection{The  nonlinear  hyperbolic oscillator: quantum case}
 
The quantum Hamiltonian $\hH$ with $\la>0$ has recently been shown to give rise to a new exactly solvable quantum model in $N$ dimensions~\cite{FTC} which has both a discrete and a continuous spectrum. The discrete spectrum depends on 
a single principal quantum number $n=0,1,2\dots$ and its eigenvalues are
\bea
  E_n \!&\!\!=\!\!&\!  -\hbar^2 \lambda\left(n + \frac{N}{2}\right)^2 + \hbar \left(n+\frac N 2\right ) \sqrt{\hbar^2 \lambda^2 \left(n+\frac{N}{2}\right)^2+ \om^2 } \nonumber\\
 \!&\!\!=\!\!&\!-\hbar^2 \lambda\left(n + \frac{N}{2}\right)^2 \left\{1 -  \sqrt{1+ \frac{\om^2 }{\hbar^2 \lambda^2 \left(n+\frac{N}{2}\right)^2}}  \right\} .
 \label{xxa}
\eea
 Therefore, the degeneracy of this model is exactly the same as in the $N$D isotropic oscillator (whose spectrum is recovered under the limit $\la\to 0$), as it should be expected from the beginning due to its MS property.
 Notice also that the bound states of this system are hydrogen-like since
$$
 \lim_{n\to \infty}E_n=\frac{\om^2}{2\la}  ,\qquad  \lim_{n\to \infty}(E_{n+1}-E_n)=0.
 $$
 Therefore 
 ${\hH}$ has an infinite number of eigenvalues contained in $(0,\frac{\om^2}{2\la} )$ and their only accumulation point is $\frac{\om^2}{2\la} $ which is,  in turn, the bottom of the continuous spectrum     given by $[\frac{\om^2}{2\la },\infty)$. The corresponding wave functions can be explicitly found in~\cite{FTC}.
 
 Finally, a short disgression on possible physical applications of this kind of exactly solvable quantum PDM Hamiltonians is in order. Firstly, we recall that the MIC-Kepler and oscillator potentials on the 3D sphere have been shown in~\cite{Gritsev} to be suitable as effective models for strong and weak confinement regimes in spherical quantum dots.  Secondly, it is worth stressing that a parabolic mass function has been proposed in~\cite{Koc,Schd} in order to describe a one-dimensional quantum well formed by a GaAs/Al$_x$Ga$_{1-x}$As heterostructure. In fact, if the concentration $x$ grows in terms of a given spatial coordinate  $q_1$ as $x(q_1)=\alpha\,q_1^2$ and in this type of material  $m^\ast(x)= m_0(a+b\, x)$~\cite{Koc}, then an effective mass function of the type $m^\ast(q_1)= m_0(a+b\, \alpha\,q_1^2)$ arises. Thus, we have obtained a realistic quantum exactly solvable model coming from the Bertrand-oscillator potential on a hyperbolic space with non-constant curvature.

\section*{Acknowledgments}

This work was partially supported by the Spanish MICINN   under grants    MTM2010-18556   and FIS2008-00209, by the    Junta de Castilla y
Le\'on  (project GR224), by the Banco Santander--UCM 
(grant GR58/08-910556)
   and by  the Italian--Spanish INFN--MICINN (pro\-ject ACI2009-1083).    
  

\end{document}